# Crystal-like Order Stabilizing Glasses: Structural Origin of Ultra-stable Metallic Glasses

Zhen Lu,[1,2] Anh Khoa Augustin Lu,[1] Fan Zhang,[2] Yuan Tian,[3] Jing Jiang,[4] Daixiu Wei,[4] Jiuhui Han,[2] Qingyang Gao,[3] Koji Ohara,[5] Hidemi Kato,[4] Akihiko Hirata,[1,2,6,†] and Mingwei Chen[3,‡]

[1]Mathematics for Advanced Materials Open Innovation Laboratory, AIST, 2-1-1 Katahira, Aobaku, Sendai 980-8577, Japan.

[2]WPI Advanced Institute for Materials Research, Tohoku University, Sendai 980-8577, Japan.

[3]Department of Materials Science and Engineering, the Johns Hopkins University, Baltimore, MD 21218, USA.

[4]Institute for Materials Research, Tohoku University, Sendai, 980-8577, Japan.

[5]JASRI/SPring-8, Sayo-cho, Sayo-gun, Hyogo 679-5198, Japan.

[6]Department of Materials Science, Waseda University, 3–4–1 Ohkubo, Shinjuku, Tokyo 169–8555, Japan.

Corresponding author

[†]ahirata@aoni.waseda.jp

[‡]mwchen@jhu.edu

## Abstract

Glasses are featured with a disordered amorphous structure, being opposite to crystals that are constituted by periodic lattices. In this study we report that the exceptional thermodynamic and kinetic stability of an ultra-stable binary ZrCu metallic glass, fabricated by high-temperature physical vapor deposition, originates from ubiquitous crystal-like medium range order (MRO) constituted by Voronoi polyhedron ordering with well-defined local translational symmetry beyond nearest atomic neighbors. The crystal-like MRO significantly improves the thermodynamic and kinetic stability of the glass, which is in opposition to the conventional wisdom that crystal-like order deteriorates the stability and forming ability of metallic glasses. This study unveils the structural origin of ultra-stable metallic glasses and shines a light on the intrinsic correlation of local atomic structure ordering with glass transition of metallic glasses.

Vitrification is conventionally achieved by rapidly cooling a liquid to bypass crystallization through a glass transition [1-3]. The dynamic transition usually occurs within a narrow temperature range where the viscosity $\eta$ of a glass-forming liquid increases extraordinarily and the resulting glass cannot adequately relax before reaching the bottom of the potential energy landscape (PEL) as an "ideal glass" with minimized entropy of amorphous structure [2,4-8]. Therefore, deep basins of the PEL are not accessible on the laboratory timescale by traditional materials processing methods, such as slow quenching, mold casting and long-time annealing [9,10]. These restrictions curb the structure and property optimization of glasses on account of kinetic and thermodynamic limitations. In contrast, by virtue of physical vapor deposition (PVD), Swallen and co-workers synthesized organic molecular glasses with prominent thermodynamic and kinetic stabilities at optimal substrate temperatures of ~0.85 $T_g$ ($T_g$ is the conventional glass transition temperatures of liquid-quenching glasses) [11-13]. These glasses, termed as ultra-stable glasses, exhibit significantly increased $T_g$, higher density and extraordinary mechanical properties [11,14-19], benefiting from a high surface molecular mobility that allows deposited molecules to efficiently sample various packing configurations for reaching deep basins in PEL [20-23]. In addition to organic glasses, ultra-stable metallic glasses (MGs), constituted by individual atoms, have also been achieved by PVD [18,24,25]. According to X-ray scattering and ellipsometry studies, ultra-stable organic glasses with a high degree of molecular complexity present anisotropically packed motifs or highly collapsed chains, as compared to their liquid-cooled counterparts [15,16,26]. However, although MGs have a relatively simple atomic structure, the structural origins of thermodynamic and kinetic stability of ultra-stable metallic glasses are largely unknown as conventional diffraction techniques lack a sufficient spatial resolution to determine local atomic configurations [27,28].

Recently-developed angstrom-beam electron diffraction (ABED) with ~3 Å coherent electron probe is a powerful tool to investigate local atomic structures of amorphous materials [29-33]. Under scanning mode, ABED generates addressable 2D diffraction mappings with an angstrom-scale spatial resolution, from which short- to medium-range structural order can be effectively characterized [34]. In this study we employed the scanning ABED technique to investigate the atomic structure of an ultra-stable $Cu_{50}Zr_{50}$ MG. Extensive crystal-like medium range order (MRO) domains, constructed by regular packing of Voronoi polyhedra with distinct local translational symmetry, were observed in the ultra-stable metallic glasses. The development of the crystal-like MRO has a strong correlation with the transition from conventional to ultra-stable glasses and is responsible to the exceptional thermodynamic and kinetic stability of the ultra-stable metallic glass.

Figure 1(a) shows the differential scanning calorimetry (DSC) curves, measured at a heating rate of 10 K/min, of the MG films with different deposition temperatures. For comparison, the glass fabricated by melt spinning, denoted as ordinary glass, is plotted as the gray curve with the lowest $T_g$ of 654 K (see Supplemental Material for details [35]). The DSC profiles of the $Cu_{50}Zr_{50}$ MG films with different deposition temperatures are scaled with the $T_g$ of the ordinary glass. Compared with the ordinary glass, the slowly grown glass films exhibit higher $T_g$ from room-temperature (RT) deposited $\Delta T_g = 23$ K to the highest value of $\Delta T_g = 46$ K at the deposition temperature of 556 K (scaled 0.85 $T_g$) [Figs. 1(a) and 1(b)]. In addition to the obviously improved $T_g$, the crystallization temperature $T_x$ also increases with deposition temperature and reaches the highest value of 737 K at the deposition temperature of 0.85 $T_g$ (denoted as the 0.85 $T_g$ ultra-stable glass) [Fig. 1(c)].

The force-depth curves of nanoindentation suggest that the ultra-stable glasses have higher deformation resistance than the ordinary one [see the inset of Fig. 1(d)]. The maximum hardness of ~ 7.997 ± 0.259 GPa is obtained from the 0.85 $T_g$ ultra-stable glass, which is obviously higher than that of the ordinary glass [7.202 ± 0.244 GPa, Fig. 1(d)]. Likewise, the 0.85 $T_g$ ultra-stable glass also presents the highest elastic modulus of 135.405 ± 3.412 GPa (see Fig. S1 [35]). The significant increments of $T_g$, $T_x$, hardness and elastic modulus confirm the previous observations that slow deposition at an optimal deposition temperature of ~0.85 $T_g$ allows the deposited atoms to bypass kinetic limitation by substantial relaxation during film growth and to reach the ultra-stable glass state [11,25].

The first diffraction peaks of X-ray diffraction (XRD) profiles of the deposited glasses slightly shift toward the high angle direction with the increase of deposition temperature, implying the increased atomic packing density at high temperatures [Fig. 2(a)]. High-resolution transmission electron microscope (HRTEM) images of the RT deposited and 0.85 $T_g$ ultra-stable glasses show homogeneous maze-like amorphous structure, together with the diffraction halos in the inserted selected area electron diffraction (SAED) patterns in Figs. 2(b) and 2(c). Obvious structure difference cannot be directly identified from the phase-contrast HRTEM images and SAED patterns. However, using mass-sensitive scanning transmission electron microscopy (STEM) with a high-angle annular dark-field (HAADF) detector, we found that the 0.85 $T_g$ ultra-stable glass presents obvious contrast variation with an autocorrelation length of $1.42 \pm 0.07$ nm [Figs. 2(d) and 2(e), see Supplemental Text 1 for details [35]]. In contrast, separate energy dispersive spectroscopy (EDS) mappings exhibit the homogeneous elements distributions in both samples down to sub-nanoscale (see Fig. S2 [35]). Since the composition distribution in the ultra-stable glass is essentially homogeneous (see Fig. S2 and Supplemental Table 1 [35]), the divergent mass contrast variation indicates the existence of atomic packing density fluctuation in the ultra-stable glass [32].

On the basis of the synchrotron X-ray diffraction spectra of PVD-deposited glasses, their pair distribution functions (PDF), $g(r)$, derived from Fourier transform of structural factors $S(Q)$, are shown in Fig. S3 [35]. The $g(r)$ of the 0.85 $T_g$ ultra-stable glass shows enhanced peak intensity up to the 4th neighboring shell, suggesting a high level of local atomic order in the ultra-stable glass. In addition, the $S(Q)$ profile of the 0.85 $T_g$ ultra-stable glass presents a unique sub-peak at ~5.3 Å$^{-1}$ (light blue box) [Fig. 2(f)], indicating an additional structure ordering in the ultra-stable glass.

To uncover the local atomic structure of the ultra-stable glasses, scanning ABED with a coherent electron beam diameter of ~3.6 Å (full width at half maximum) is employed to characterize RT-deposited and 0.85 $T_g$ ultra-stable glasses from a large area of 8 nm by 8 nm with a pixel size of 0.2 nm × 0.2 nm [Fig. 3(a), Fig. S5]. On account of the unique diffraction peak at 5.3 Å$^{-1}$ of the 0.85 $T_g$ ultra-stable glass [Fig. 2(f)], we selected the diffraction vector with the length from 5.2 Å$^{-1}$ to 5.4 Å$^{-1}$ (Fig. S5), and constructed the 2D ABED mappings (6400 ABED patterns for each mapping) as shown in Figs. 3(b) and 3(c). The brighter regions represent the ordering domains that have a scattering vector of ~5.3 Å$^{-1}$ from the unique structure ordering of the 0.85 $T_g$ ultra-stable glass [Fig. 2(f), see Supplemental Text 2 for details [35]]. In comparison with the RT-deposited sample, the 0.85 $T_g$ ultra-stable glass exhibits relatively large ordering domains with a size of ~2 nm. Interestingly, the size and morphology of the domains are coincident with these of the spatial heterogeneity revealed by the HAADF image of Fig. 2(e). Therefore, both real- and reciprocal-space results suggest that a pronounced local structure ordering beyond SRO takes place in the 0.85 $T_g$ ultra-stable glass during the high-temperature and slow deposition.

To further expose the structural signature of the local ordering in the ultra-stable glass, ABED diffraction patterns from the bright regions in Figs. 3(b) and 3(c) are analyzed. Two kinds of ABED patterns with symmetric diffraction spots are frequently observed [Figs. 3(d) and 3(e)]. In particular, the second-order diffraction spots with a scattering vector of ~5.4 Å$^{-1}$ indicate that weak yet detectable translational (i.e., crystal-like) symmetry may exist in these locally ordered domains. Reverse Monte Carlo (RMC) simulations based on the initial structure constructed by MD modeling of $Cu_{50}Zr_{50}$ and subsequent structure analysis of individual polyhedra were performed to build up a reference database for determining the corresponding atomic structures of these ABED patterns (Fig. S6 and Fig. S8, see Supplemental Text 3 for details [35]). Since the crystal-like structure manifests the decreased number of five-edge pentagonal faces of the Cu and Zr centered polyhedra, which is reflected by the third digit in the Voronoi indices [42]. We thus simulated $S(Q)$ profiles of the representative single polyhedra with a relative low number of pentagons (Fig. S9 and Fig. S10 [35]). We found that Cu-centered <0,3,6,3> and Zr-centered <0,4,4,6> polyhedra present strong diffraction intensity at ~5.3 Å$^{-1}$ (Fig. S10 [35]). The simulated ABED patterns of <0,3,6,3> and <0,4,4,6> polyhedra [Figs. 4(b) and 4(f)] well reproduce the experimental ones in terms of diffraction vectors (lengths and angles), e.g., the vector ratio $d_1/d_2$ gives the values of 1.09 ± 0.05 and 1.06 ± 0.2 for Figs. 4(a) and 4(b), 1.04 ± 0.3 and 1.07 ± 0.2 for Figs. 4(e) and 4(f). Since the <0,4,4,6> and <0,3,6,3> polyhedra can easily transform into the bcc polyhedron of <0,6,0,8> by slightly changing the positions of three atoms as illustrated in Figs. 4(c), 4(d) and 4(g), 4(h) [43,44], they can be considered as distorted bcc-like polyhedral topologies. In fact, the ABED patterns of these <0,4,4,6> and <0,3,6,3> polyhedra are very similar to those of the bcc <0,6,0,8> polyhedron along a specific orientation, for example, the <0,3,6,3> ABED patterns in Figs. 4(e) and 4(f) are analogous to that of the bcc-like polyhedron along [111] direction [Fig. S(11) [35]]. The rich four-fold and six-fold faces of <0,4,4,6> and <0,3,6,3> polyhedra make it possible to form the cooperatively arranged amorphous polyhedra by local translational symmetry operations.

Figure 4(i) shows 28 ABED patterns taken consecutively with a scan step of 2.0 Å from an 8 nm by 8 nm area of the 0.85 $T_g$ ultra-stable glass. The diffraction vectors marked with white arrows on left-lower side of the ABED patterns denote the diffraction spot near 5.3 Å$^{-1}$. All the ABED patterns in the highlighted region can be indexed as <0,4,4,6> polyhedra. Importantly, these electron diffraction patterns have nearly the same orientation, demonstrating that the corresponding <0,4,4,6> polyhedra are packed together with parallel polyhedral faces and constitute an MRO domain by local translational symmetry operations. The crystal-like MRO spans ~1.6 nm along the longest direction, which is consistent with the dimensions of densely packed domains revealed by HAADF-STEM [Fig. 2(e)]. Similarly, crystal-like MRO domains constituted by <0,3,6,3> polyhedra can also be observed in the 0.85 $T_g$ ultra-stable glass (Fig. S12 [35]). Although distorted icosahedral polyhedra (<0,0,12,0> or <0,2,8,2>) are also frequently observed in the blue regions [Figs. 3(b) and 3(c)], they usually appear as SRO clusters and do not form MRO domains alone, which is in line with the previous observations in hard spheres glasses [45]. Therefore, the crystal-like MRO with distinct translational symmetry appears to be the structural origin of the enhanced thermal and kinetic stabilities of the ultra-stable metallic glass. Importantly, although the polyhedral clusters are packed by the translational symmetry operations, individual building blocks, i.e., polyhedral atomic clusters, are not geometrically and chemically identical. Consequently, the crystal-like packing scheme of polyhedra cannot generate atomic lattices with well-defined periodicity and interatomic distances and, in fact, the MRO is hidden in overall structure disorder when individual atoms are counted as the structural components. However, with the overall structural disorder, the formation of crystal-like packing of few low-energy, densely-packed polyhedra definitely leads to the decrease of configurational entropy of the ultra-stable, indicating a possible packing scheme of "ideal glasses" in atomic and hard-sphere systems [4,8,23,46].

Although crystal-like local order has been proposed to describe the structures of conventional glasses in terms of packing efficiency [32,47-55], the crystal-like MRO domains with a size of 1-2 nm, constituted by amorphous polyhedron atomic clusters, have not been experimentally observed before and represent an important structural feature of the ultra-stable metallic glass. As the conventional wisdom, crystal-like local structures can act as the nuclei of thermodynamically stable crystalline phases and deteriorate the thermal and kinetic stability of glasses. Therefore, the finding of extensive crystal-like MRO in the ultra-stable metallic glass provides a new insight into the intrinsic correlation of thermodynamic and kinetic stability of MGs with their local translational structure ordering. Since the ultra-stable glass is prepared *via* slow vapor deposition at high temperatures, deposited atoms and atomic clusters have sufficient time to arrange themselves by forming low-energy icosahedron-like or crystal-like atomic configurations [56]. The lack of translational symmetry prevents the growth of icosahedron-like order from SRO to MRO whereas the crystal-like order can grow [45]. However, limited by the geometrical frustration arising from the competition between the large atomic size difference and the negative enthalpy of mixing between Zr and Cu [30,50,57,58], the crystal-like domains with a large elastic distortion and insufficient local translational symmetry of individual amorphous polyhedra cannot develop to larger than the critical nucleation size of a crystalline phase and thus remain to be MRO. According to the classical nucleation theory and the Gibbs-Thomson equation [59,60], the calculated critical nucleation size of Cu$_{50}$Zr$_{50}$ at 700 K ($T_g$ of ultra-stable glass) is ~2.88 nm [see Supplemental Text 5 for details [61]]. Apparently, the low-energy crystal-like MRO domains with the size around 1-2 nm [Fig. 4(i) and Fig. S12 [35]] are thermodynamically and kinetically stable and cannot evolve as the crystal nuclei at the deposition temperature of $0.85T_g$. Since the critical nucleation size for crystallization increases with the decrease of undercooling, i.e., $T_\infty - T_d$, the crystal-like MRO thus contributes to outstanding thermal stability of ultra-stable glass with a higher crystallization temperature. Moreover, the formation of these crystal-like MRO domains leads to spatial heterogeneity [Fig. 2(e)] [32], which are expected to generate multiple occupied energy states in PEL. Especially, these at the deep basins require higher kinetic energy and thereby higher temperatures to activate the glassy system into a supercooled liquid state [50]. Consequently, the ultra-stable glass with prominent crystal-like order possesses a higher $T_g$ for the kinetic transition from a glass to a supercooled liquid.

**Acknowledgements**

We thank the Common Equipment Unit in AIMR, Tohoku University for the materials analysis facilities. S. C. Ning, F. Zhu and K. Nishio are thanked for discussions. This work was sponsored by the Fusion Research Funds from WPI-AIMR, Tohoku University. M.C. is supported by U.S. National Science Foundation under grant DMR-1804320.

**Figure Captions**

FiG. 1. (a) Representative DSC heat flow curves of $Cu_{50}Zr_{50}$ MGs at a heating rate of 10 K/min. $T_g$ and $T_x$ are defined from the intersection of black lines of the onsets of transformations. (b) and (c) $T_g$ and $T_x$ of $Cu_{50}Zr_{50}$ MGs, respectively (the black dash lines are a guide for the eye). The error bars are given by three measurements. (d) The hardness of $Cu_{50}Zr_{50}$ MGs and the inset shows the corresponding force-depth curves.

FIG. 2. (a) XRD profiles of $Cu_{50}Zr_{50}$ MGs films confirming their amorphous nature. (b) and (c) HRTEM images of RT-deposited and 0.85 $T_g$ ultra-stable glasses and corresponding SAED patterns (inset). Scale bar: 5 nm. (d) and (e) HAADF-STEM images of RT-deposited and $0.85T_g$ ultra-stable glasses, respectively. Scale bar: 5 nm. (f) The magnified $S(Q)$ diffraction profiles of RT-deposited and 0.85 $T_g$ ultra-stable glasses. Inset: the whole $S(Q)$ curves for two glasses.

FIG. 3. (a) Experimental procedure of the scanning ABED experiments with an electron beam diameter of ~3.6 Å. (b) and (c) Reconstructed 2D ABED diffraction mappings with selected diffraction vector lengths between 5.2 Å$^{-1}$ to 5.4 Å$^{-1}$ of RT-deposited and $0.85T_g$ ultra-stable glasses, respectively. Scale bar: 2 nm. (d) and (e) Frequently observed ABED patterns containing six distinguishable diffraction spots and strong diffraction spots near 5.3 Å$^{-1}$.

FIG. 4. (a) and (e) Experimental featured ABED patterns of the ultra-stable MG; and (b) and (f) simulated ABED patterns from <0,4,4,6> and <0,3,6,3> polyhedra, respectively. (c) and (d), (g) and (h) Schematics of the transformation of <0,4,4,6> and <0,3,6,3> into the bcc <0,6,0,8> polyhedron, respectively. (i) Twenty-eight ABED patterns taken consecutively from Fig. 3(c) from a bright region showing a crystal-like MRO domain.

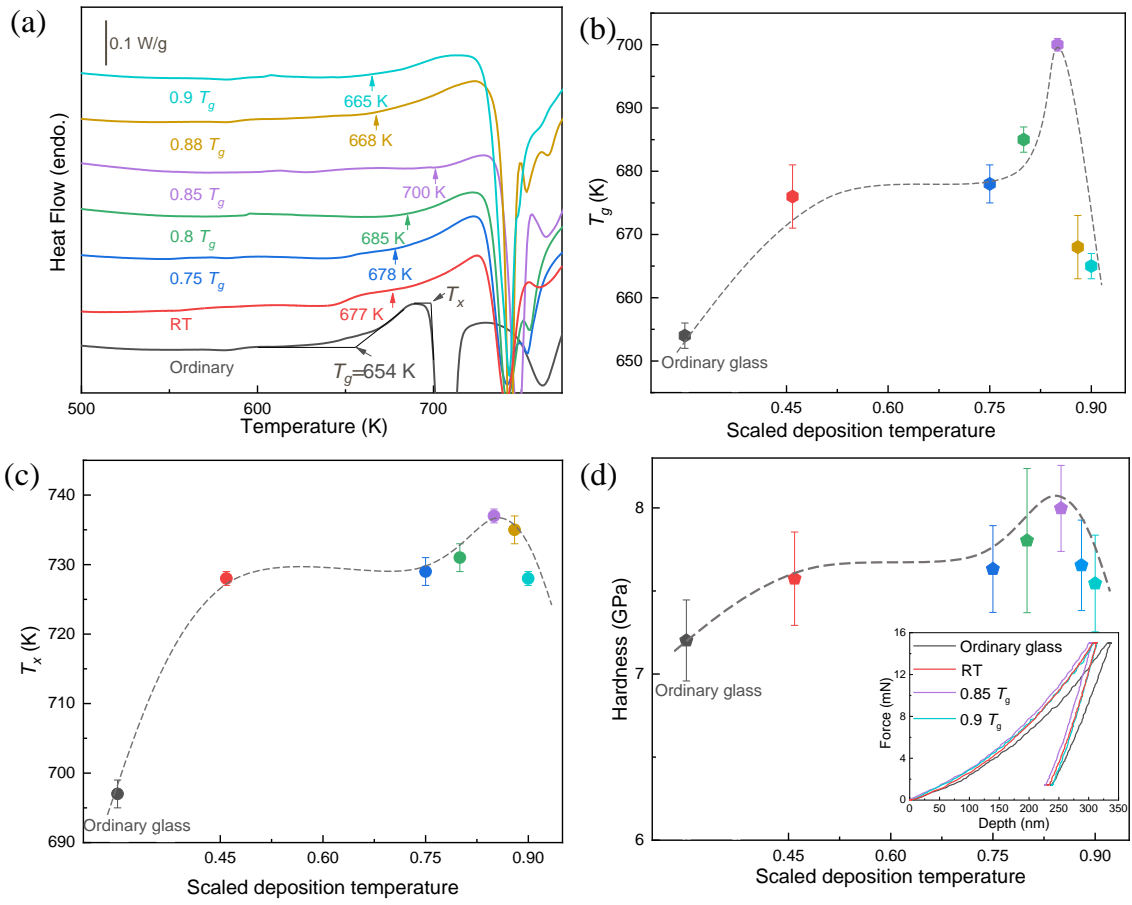

**Fig. 1. Lu** *et al.*

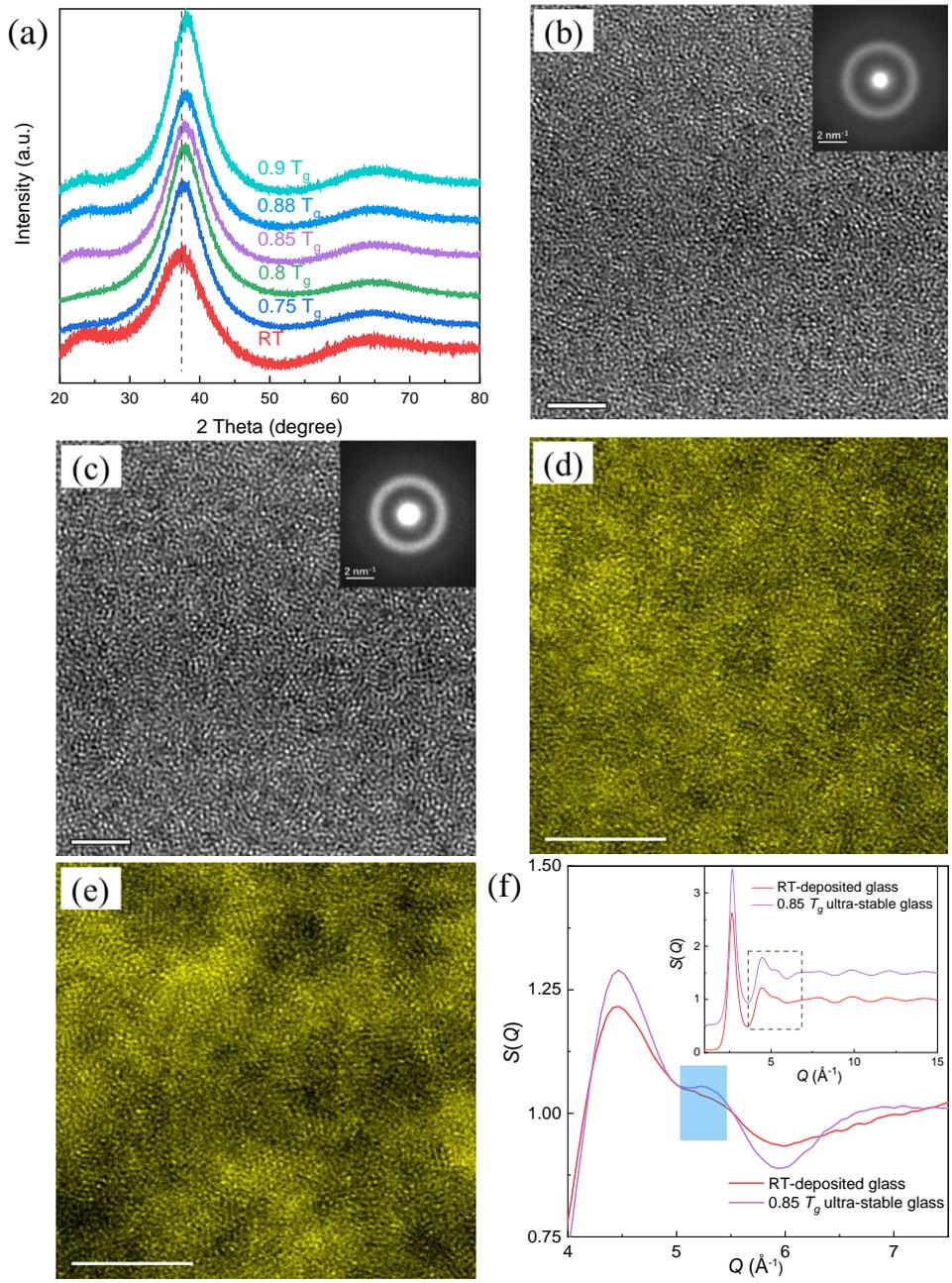

**Fig. 2. Lu** *et al.*

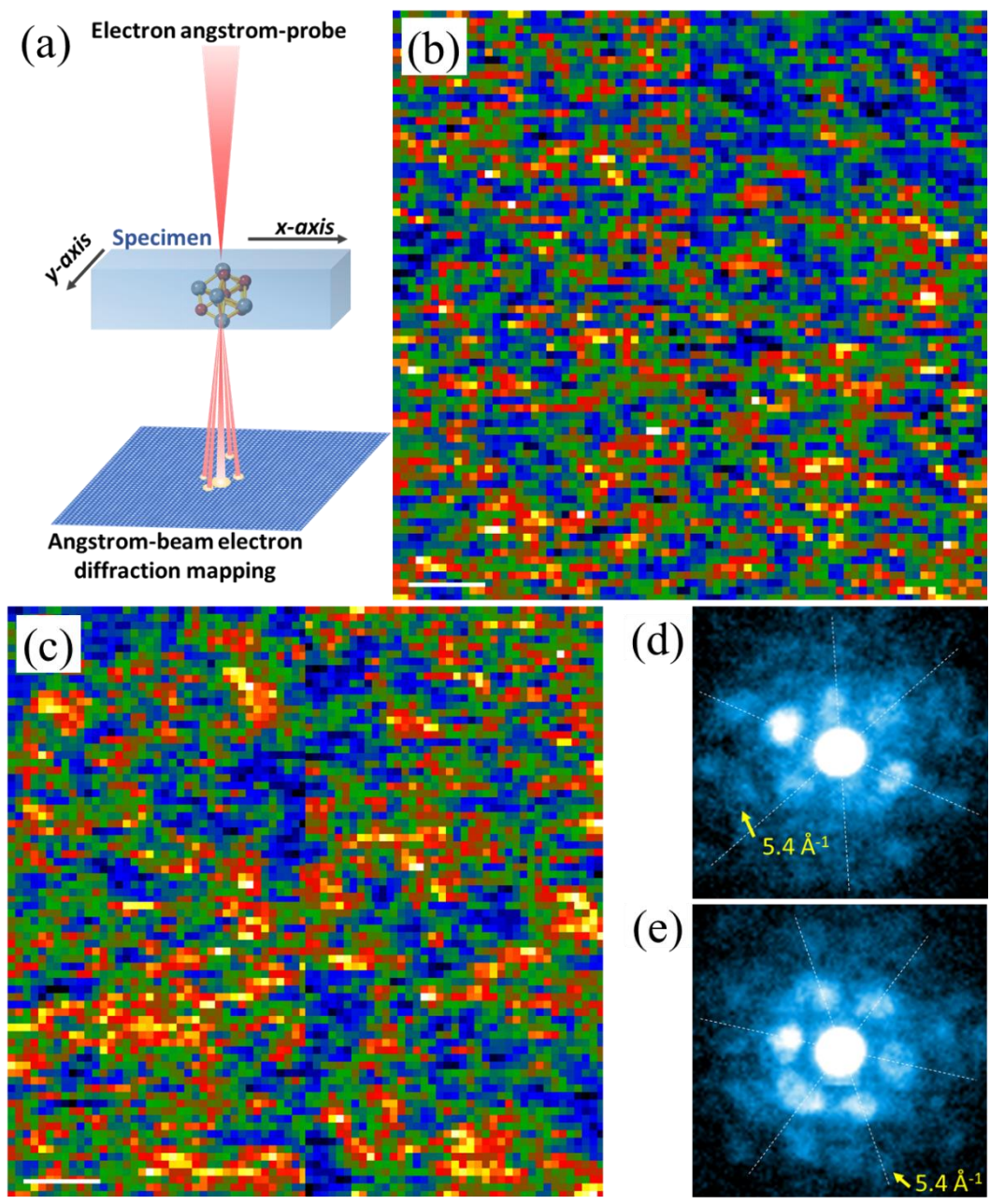

Fig. 3. Lu *et al.*

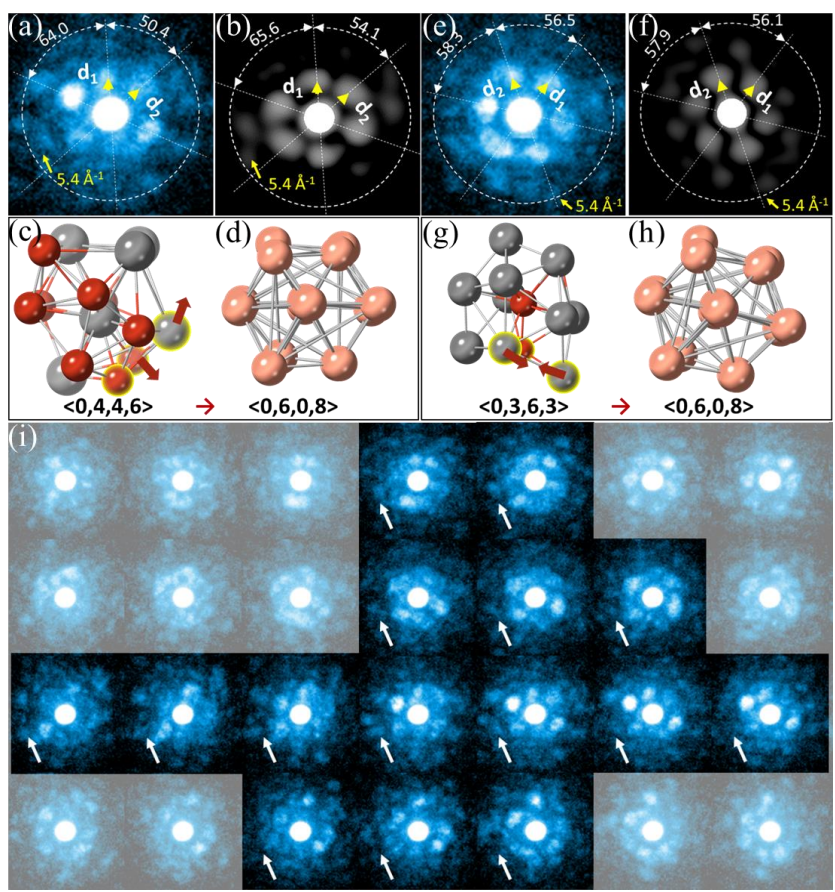

**Fig. 4. Lu** *et al.*